\documentclass[pra, twocolumn, floatfix, superscriptaddress, longbibliography, preprintnumbers]{revtex4-1}

\usepackage{amsmath, amssymb, bbm, bbold}
\usepackage{graphicx}
\usepackage{color}
\usepackage{hyperref}

\definecolor{dgreen}{rgb}{0,1,1}

\begin{document}

\title{Eigenstate extraction with neural-network tomography}

\author{Abhijeet Melkani}
\email{amelkani@uoregon.edu; present address: Department of Physics, University of Oregon, Eugene, OR-97405, USA}
\affiliation{Theoretical Quantum Physics Laboratory, RIKEN Cluster for Pioneering Research, Wako-shi, Saitama 351-0198, Japan}
\affiliation{Department of Physics, Indian Institute of Technology Bombay, Mumbai, Maharashtra 400076, India}

\author{Clemens Gneiting}
\email{clemens.gneiting@riken.jp}
\affiliation{Theoretical Quantum Physics Laboratory, RIKEN Cluster for Pioneering Research, Wako-shi, Saitama 351-0198, Japan}

\author{Franco Nori}
\affiliation{Theoretical Quantum Physics Laboratory, RIKEN Cluster for Pioneering Research, Wako-shi, Saitama 351-0198, Japan}
\affiliation{Department of Physics, University of Michigan, Ann Arbor, Michigan 48109-1040, USA}

\date{\today}

\begin{abstract}
We discuss quantum state tomography via a stepwise reconstruction of the eigenstates of the mixed states produced in experiments. Our method is tailored to the experimentally relevant class of nearly pure states, or simple mixed states, which exhibit dominant eigenstates and thus lend themselves to low-rank approximations. The developed scheme is applicable to any pure-state tomography method, promoting it to mixed-state tomography. Here, we demonstrate it with machine learning-inspired pure-state tomography based on neural-network representations of quantum states. The latter have been shown to efficiently approximate generic classes of complex (pure) states of large quantum systems. We test our method by applying it to experimental data from trapped ion experiments with four to eight qubits.
\end{abstract}

\preprint{\textsf{published in Phys.~Rev.~A~{\bf 102}, 022412 (2020)}}

\maketitle

\section{Introduction}

In times where quantum experiments and quantum devices have reached unprecedented size and complexity, their verification has become increasingly hard and yet indispensable. Noise and imperfections cause deviations of the produced states from the target states, which may, in many cases, put their intended purpose in jeopardy. Quantum state tomography is the process of reconstructing the states produced in quantum experiments or devices from their measurement data. Based on statistical analysis of a near-complete set of this measurement data, the realized states can, in principle, be fully reconstructed with high accuracy \cite{Paris2004quantum, Hayashi2005asymptotic, Lvovsky2009continuous, Hou2016full}.

However, full, unconditional quantum state tomography becomes prohibitively expensive with increasing Hilbert space dimension, both from an experimental perspective (the required number of measurements scales exponentially with the system size) and from the perspective of data post-processing. Strategies to mitigate these costs include exploiting symmetries, minimizing the number of required measurements, or adaptive measurement schemes \cite{Gross2010quantum, Cramer2010efficient, Toth2010permutationally, Hou2016achieving, Kaznady2009numerical}.

One reason behind this cost explosion is that full state tomography recovers the entire quantum state, while, in many circumstances, low-rank approximations are sufficient to retrieve the relevant information. This is, in particular, the case if the target state is pure and the produced state thus can be expected to exhibit a clear hierarchy in its spectrum, featuring a dominant eigenvalue/eigenstate pair, followed by increasingly irrelevant subdominant eigenvalue/eigenstate pairs.

\begin{figure}[htb]
\includegraphics[width=0.9\columnwidth]{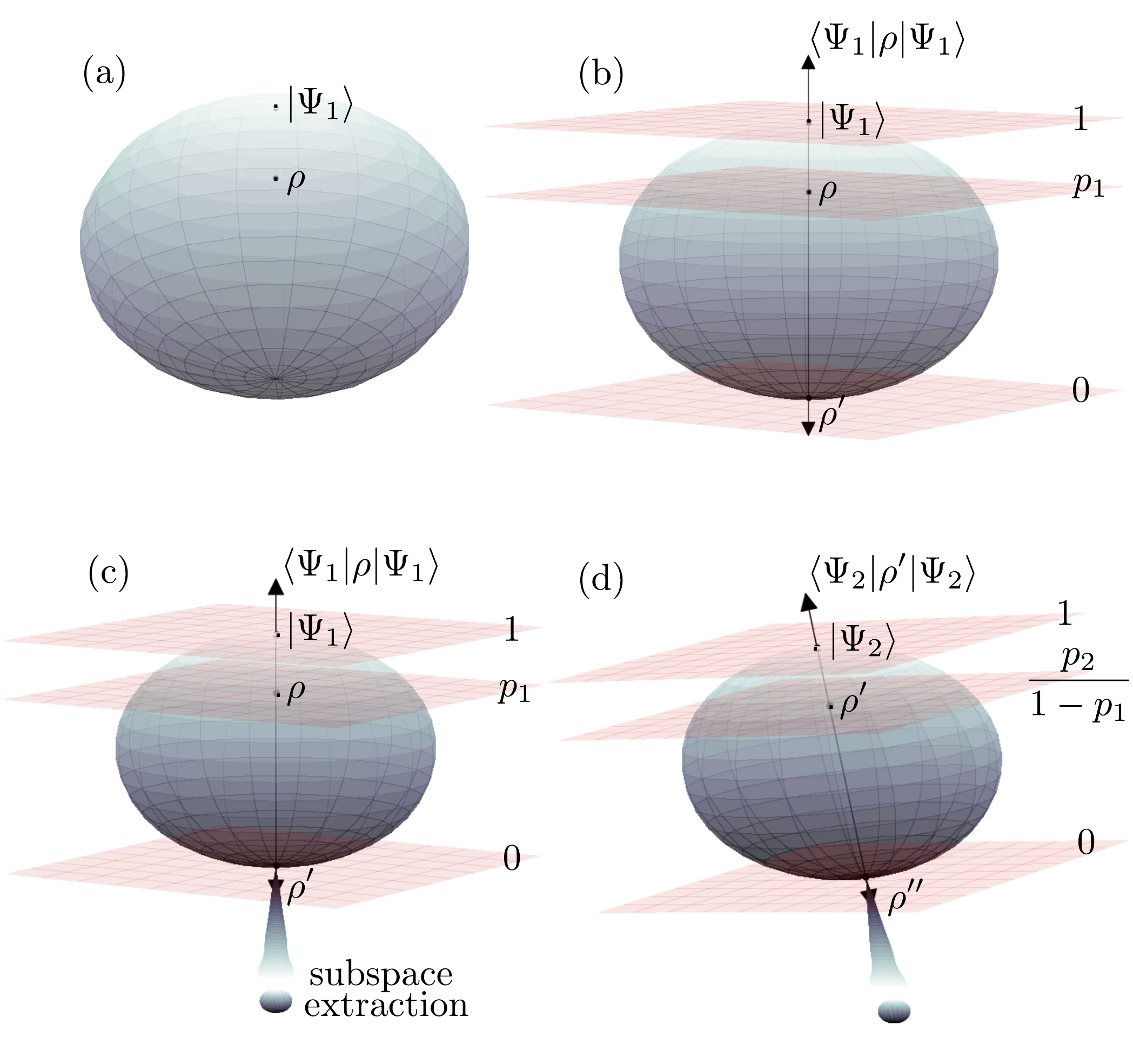}
	\caption{\label{Fig:Eigenstate_tomography} Stepwise reconstruction of a (nearly pure) mixed state $\rho = \sum_{i=1}^{n} p_i |\Psi_i\rangle \langle\Psi_i|$ ($p_{i+1} \leq p_i$) from measurement data. (a) The mixed state $\rho$ resides in the $(N^2 - 1)$-dimensional, convex space $M(N)$ of $N$-dimensional density matrices. The $(N^2 - 2)$-dimensional boundary of $M(N)$ is composed of all density matrices with rank $r<N$, and the pure states form a continuous, $2(N - 1)$-dimensional subset of this boundary \cite{bengtsson_zyczkowski_2017}. The closest (in terms of trace distance) pure state to $\rho$ is its dominant eigenstate $|\Psi_1\rangle$. (b) A neural-network ansatz, constrained to the submanifold of pure states, is trained to find the closest pure state in compliance with the measurements. It thereby approximates the dominant eigenstate $|\Psi_1\rangle$ and its corresponding eigenvalue, the dominant eigenvalue $p_1$, which captures the distance of $|\Psi_1\rangle$ from $\rho$. Geometrically, this corresponds to finding the direction of projection along which $\rho$ is closest to the boundary. This projection is equivalent to the operation $\langle\Psi_1|\rho|\Psi_1\rangle$, mapping $M(N)$ to a set of hyperplanes corresponding to different values of $\langle\Psi_1|\rho|\Psi_1\rangle$. (c) The state $\rho'$, defined by $(1 - p_1)\rho' = \rho - p_1|\Psi_1\rangle\langle \Psi_1|$, belongs to the $[(N-1)^2 - 1]$-dimensional subspace of density matrices formed by the intersection of $M(N)$ and the ``0-hyperplane''. This subspace is ``extracted'' (shown schematically) and (d) constitutes the starting point for the next iteration step based on $\rho'$. Repeating this procedure then allows one to collect the eigenstate/eigenvalue pairs of $\rho$ up to desired rank.}
\end{figure}

In this paper, we leverage on this idea, proposing the stepwise reconstruction of quantum states in terms of their leading eigenvalue/eigenstate pairs, cf.~Fig.~\ref{Fig:Eigenstate_tomography}. Our scheme is based on the insight that highly efficient methods for pure-state tomography can also be used to robustly recover the (dominant) eigenstates of mixed states. Tailored iteration then allows one to recover the eigenvalue/eigenstate pairs of mixed states up to a desired rank. Such (low-rank) reconstruction of mixed states not only delivers valuable structural information about the state produced but also comes with substantially reduced costs.

Our scheme promotes any pure-state tomography method, i.e., any measurement data-based state estimation that is constrained within the set of pure states, to mixed-state tomography. Such restriction to pure states can be favorable for several reasons. Besides being computationally more efficient, many physically motivated many-body ansatz states, e.g., matrix product states, entangled plaquette states, and string-bond states, are naturally formulated in terms of pure states \cite{Glasser2018neural}. Moreover pure states are conceptually simpler \cite{xin2020hybrid_quantumclassical, huang2017math_pure} and thus allow tailored approaches such as, e.g., disentangling the state at local sites \cite{Cramer2010efficient}, or reconstructing the generating unitary \cite{xin2020hybrid_quantumclassical}.

To demonstrate our reconstruction scheme, we here use and adjust a recently developed method for pure-state tomography based on neural-network representations of quantum states \cite{Torlai2018neural}. Neural network representations have been proven versatile in an increasing number of applications in quantum physics \cite{Carleo2017solving, Carrasquilla2017machine, Deng2017quantum, Glasser2018neural, Venderley2018machine}, and quantum state tomography appears particularly well-suited, due to its inherently data-driven nature. Indeed, {\it Neural Quantum States} (NQS) have been shown \cite{Torlai2018neural} to be viable for tomography of complex, high-dimensional pure states, leveraging both the efficient and scalable representation of neural networks and their great expressional power. As we show here, these benefits carry over to the eigenstate reconstruction of mixed states.

Our reconstruction scheme may be reminiscent of {\it Principal Component Analysis} (PCA), a common technique in data science, which also aims at approximating matrices in terms of their spectral properties. However, in contrast to PCA, where {\it a priori} knowledge of the full matrix is presumed, here, the eigenvalue/eigenstate pairs are iteratively reconstructed, directly from the measurement data, and only up to a desired rank. In this sense, our method represents a systematic way to reconstruct the density matrix step-by-step, targeting states which exhibit a clear hierarchy among their eigenvalues (i.e., states with low entropy). In contrast to other methods based on low-rank approximations (e.g., Ref.~\cite{Gross2010quantum}), our scheme does not require to specify {\it a priori} the rank of the approximation. Moreover, in the presence of generic noise, the proposed, stepwise low-rank reconstruction may deliver a more faithful reconstruction of the produced state, as compared to extracting this information from single-shot low-rank approximations.

This article is structured as follows: In Section~\ref{Sec:Theoretical_Results} we develop the theoretical foundation towards the robust recovery of the dominant eigenstates. We then introduce our iterative eigenstate reconstruction method in Section~\ref{Sec:Methods}. Section~\ref{Sec:Neural_quantum_states} reviews the Neural Quantum States and their utilization for pure-state and mixed-state tomography \cite{Carleo2017solving, Torlai2018neural}. In Section~\ref{Sec:Mixed_state_tomography}, we then demonstrate our iterative reconstruction method, based on neural-network pure-state tomography, using experimental data from trapped-ions experiments. Finally, we summarize our results, along with an outlook, in the Conclusions.

\section{\label{Sec:Theoretical_Results} Optimal low-rank approximations of density matrices}

We begin with formulating four propositions on low-rank approximations, which provide us with the theoretical underpinning for the reconstruction of mixed states from pure-state tomography. Their proofs are detailed in Appendix~\ref{Sec:App-Theory}.

Let us assume $\rho$ to be the density matrix to be reconstructed,
\begin{equation}\label{rho_definition}
	\rho = \sum_{i=1}^n p_i |\Psi_i\rangle\langle\Psi_i| ,
\end{equation}
where $p_1 \geq p_2 \geq ... \geq p_n$. As the two most relevant measures of distance between quantum states, we consider the fidelity $F$ between two states $\rho$ and $\sigma$,
\begin{equation}\label{Fidelity-Definition}
	F(\rho, \sigma) = \bigg [\text{Tr}\bigg (\sqrt{\sqrt{\sigma} \rho \sqrt{\sigma}}\bigg )\bigg ]^2 ,
\end{equation}
and their trace distance $T$,
\begin{align}\label{TD-Definition}
	T(\rho, \sigma) &= \frac{1}{2}\text{Tr}\sqrt{(\rho - \sigma)^{\dagger}(\rho - \sigma)} \\
	&= \frac{1}{2}\text{Tr}|\rho - \sigma| = \frac{1}{2}\sum_{i=1}^n|\lambda_i| , \nonumber
\end{align}
where $\lambda_i$ are the eigenvalues of the Hermitian matrix $(\rho - \sigma)$. We then find the following propositions regulating the recovery of the dominant eigenstates of the density matrix $\rho$:

\emph{Proposition 1:} In terms of fidelity, the unique closest pure state to $\rho$ is its dominant eigenstate $|\Psi_1\rangle$, with fidelity $F(\rho, |\Psi_1\rangle \langle \Psi_1|) = p_1$.

\emph{Proposition 2:} In terms of trace distance, the unique closest pure state to $\rho$ is its dominant eigenstate $|\Psi_1\rangle$, with trace distance $T(\rho, |\Psi_1\rangle \langle \Psi_1|) = 1 - p_1$.

\emph{Proposition 3:} In terms of fidelity, the unique closest rank-$r$ approximation to $\rho$ is
\begin{equation}\label{sigma_definition}
	\sigma = \kappa(r)^{-1} \sum_{i=1}^r p_i |\Psi_i\rangle\langle\Psi_i| ,
\end{equation}
with fidelity $F(\rho,\sigma) = \kappa(r)$ and $\kappa(r) := p_1 + p_2 + ... + p_r$.

\emph{Proposition 4:} There are infinitely many rank-$r$ approximations to $\rho$ which achieve the same trace distance as $\sigma$, $T(\rho, \sigma) = 1 -\kappa(r)$.

Detailed proofs of these propositions can be found in Appendix~\ref{Sec:App-Theory}.

We can draw several conclusions from these propositions: First, Propositions 1 and 2 clarify that pure-state approximations can reliably recover the dominant eigenstate of a density matrix. Notably, such reconstruction appears robust under variation of the underlying distance measure. This is important, because in tomography, distance measures can only be approximately reconstructed from the finite measurement data. On the other hand, Propositions 3 and 4 indicate that single-shot rank-$r$ approximations are prone to degeneracies, depending on the choice of distance measure, and thus putting the successful reconstruction of the state in jeopardy.

These conclusions motivate an iterative approach for state recovery, based on stepwise pure-state approximations. An algorithm for the iterative reconstruction of a rank-$r$ approximation will be outlined in the following section.

We remark that, in the context of tomography, where explicit representations of target states are not available, the calculation of fidelity typically scales exponentially, which renders it intractable even for moderate numbers of qubits \cite{Learnability_scaling}. In that case, one may confine to local observables as estimators of accuracy (which can be sampled efficiently from the restricted Boltzmann machines outlined below \cite{Learnability_scaling}). Here again, iterative pure-state approximations appear preferrable to single-shot rank-$r$ approximations, due to their inherent robustness with respect to distance measures.

\section{\label{Sec:Methods} Iterative eigenstate reconstruction}

We now present a scheme which, in principle, promotes any method for pure-state tomography to mixed-state tomography. Below, we will demonstrate this via pure-state tomography with neural-network quantum states \cite{Torlai2018neural}. In the following, let $P_m$ denote a family of projectors, corresponding to measurements in the experiment.

\subsection*{Schematic mixed-state reconstruction}

\emph{STEP 1:} Based on the measurement statistics $\text{Tr}(P_m\rho)$, employ a chosen method for pure-state tomography to determine the pure state $|\Hat{\Psi}_1\rangle$ which is closest (by a chosen distance measure) to $\rho$.\\

\emph{STEP 2:} Numerically calculate the measurement statistics for the eigenstate approximation $|\Hat{\Psi}_1\rangle$, $\text{Tr}(P_m|\Hat{\Psi}_1\rangle \langle \Hat{\Psi}_1|) = \langle \Hat{\Psi}_1|P_m|\Hat{\Psi}_1\rangle$.\\

\emph{STEP 3:} Determine the dominant eigenvalue $\hat{p}_1$ corresponding to $|\Hat{\Psi}_1\rangle$  (the procedure for which is discussed below). Then calculate the measurement statistics for the hypothetical state $\rho'$ according to
\begin{equation}\label{newMeas}
\text{Tr}(P_m\rho') = \frac{1}{1-\hat{p}_1} \big( \text{Tr}(P_m\rho) - \hat{p}_1\langle \Hat{\Psi}_1|P_m|\Hat{\Psi}_1\rangle \big) ,
\end{equation}\\
where
\begin{align} \label{rho_Dash_Def}
	\rho' &= \frac{1}{1-\hat{p}_1}(\rho-\hat{p}_1|\Hat{\Psi}_1\rangle\langle \Hat{\Psi}_1|) \nonumber \\
	&\approx \frac{1}{1-\hat{p}_1}\sum_{i=2}^n p_i|\Psi_i\rangle\langle\Psi_i|
\end{align}
describes the unknown state $\rho$ reduced by its dominant eigenstate contribution. Note that the estimation of $\hat{p}_1$ should guarantee that the resulting measurement probabilities remain well-defined, i.e., nonnegative. The hypothetical state $\rho'$, on the other hand, is merely an auxiliary construct and thus is not required to be physical.

\emph{STEP 4:} Return to STEP 1 with the new measurement statistics $\text{Tr}(P_m\rho')$, and proceed to extract $|\Hat{\Psi}_2\rangle$ and $\hat{p}_2'=\hat{p}_2/(1-\hat{p}_1)$, i.e., the new largest eigenstate/eigenvalue pair.\\

\emph{TERMINATION:} Terminate after reaching the desired rank $k$, i.e., after extracting the $k$ eigenstates corresponding to the first $k$ largest eigenvalues. The constructed (normalized) density matrix then becomes
\begin{equation}\label{Rho_Estimated}
	\hat{\rho} = \frac{1}{\sum_{i=1}^k\hat{p}_i} \sum_{i=1}^k \hat{p}_i|\Hat{\Psi}_i\rangle\langle\Hat{\Psi}_i| .
\end{equation}
We emphasize that, by construction, the scheme delivers the spectral decomposition of well-defined quantum states. This is in particular the case, if, as we implement it, subsequent eigenstate approximations are forced to be orthogonal to the preceding ones. The scheme can be expected to provide us with accurate state approximations up to rank $r$, if $||\hat{p}_i |\hat{\Psi}_i\rangle - p_i |\Psi_i\rangle|| \ll p_{r+1}$ for all $i \le r$. 

To verify that an iteration step results in an improved state approximation, one can, for instance, compare the respective (before and after the additional iteration step) likelihood functions of the state approximations with respect to the measurement data. If the likelihood improves, then the step is approved and a further iteration step can be tried (if desired). If the likelihood remains unchanged or deteriorates, then the step is discarded and the iteration terminates.

\subsection*{Estimation of the dominant eigenvalue}

Knowledge of the approximation $|\Hat{\Psi}_1\rangle$ to the dominant eigenstate $|\Psi_1\rangle$ allows us to estimate the corresponding eigenvalue $p_1$ from the measurements statistics $\text{Tr}(P_m\rho)$. In principle, the trace distance provides us with a straightforward way to retrieve a corresponding estimation $\hat{p}_1$, since
\begin{equation}\label{TD_And_P1}
	1 - p_1 = T(\rho, |\Psi_1\rangle\langle\Psi_1|) .
\end{equation}
The trace distance, in turn, can be estimated from the measurement statistics according to
\begin{equation}\label{TD_And_Projectors}
	T(\rho, |\Psi_1\rangle\langle\Psi_1|) = \text{max}_{P} |\text{Tr}(P\rho) - \langle \Psi_1|P|\Psi_1\rangle| ,
\end{equation}
where the maximization is over all projectors \cite{Nielsen2011QCQ}. We then estimate
\begin{align}\label{TD_And_ProjectorsM}
	1 - \hat{p}_1 &\approx 1 - p_1\nonumber
	\\ &= T(\rho, |\Psi_1\rangle\langle\Psi_1|) \nonumber \\
	&\approx \text{max}_{P_m} |\text{Tr}(P_m\rho) - \langle \Psi_1|P_m|\Psi_1\rangle| ,
\end{align}
where the $P_m$ denote the measurement projectors used in the tomography experiment. However, the naive application of Eq.~(\ref{TD_And_ProjectorsM}) is problematic, since it systematically overestimates $p_1$, which then results in unphysical measurement statistics in Eq.~(\ref{newMeas}). The reason for the overestimation is that the finite set of measurement operators $P_m$ is unlikely to contain the (close to) maximizing projectors.

To exclude unphysical measurement statistics, we here choose to employ the estimate
\begin{equation}\label{PHat-Def}
	p_1^b = \text{min}_{P_m} \frac{\text{Tr}(P_m\rho)}{\langle \Hat{\Psi}_1|P_m|\Hat{\Psi}_1\rangle} ,
\end{equation}
which follows from the constraint $\text{Tr}(P_m\rho') \ge 0$, cf.~Eq.~(\ref{newMeas}). Using $p_1^b$ as $\hat{p}_1$ thus guarantees by construction that the subsequent measurement statistics remain nonnegative. Note that the estimate is exact if both (i) $|\Hat{\Psi}_1\rangle = |\Psi_1\rangle$ and (ii) there exists an $m$, such that $P_m = |\Psi_1\rangle \langle \Psi_1|$, as can easily be seen by decomposing
\begin{equation}\label{split_P1B}
p_1^b = \text{min}_{P_m} \left( p_1\frac{\langle \Psi_1|P_m|\Psi_1\rangle}{\langle \Hat{\Psi}_1|P_m|\Hat{\Psi}_1\rangle} + \sum_{i=2}^np_i\frac{\langle \Psi_i|P_m|\Psi_i\rangle}{\langle \Hat{\Psi}_1|P_m|\Hat{\Psi}_1\rangle} \right) .
\end{equation}
We remark that, if $|\Hat{\Psi}_1\rangle = |\Psi_1\rangle$, then $p_1^b \leq 1$ cannot underestimate $p_1$. On the other hand, if $|\Hat{\Psi}_1\rangle \neq |\Psi_1\rangle$, then we find numerically that the measurements ${P_m}$, which underestimate $p_1$ (and thus determine $p_1^b$), have little overlap with $|\Psi_1 \rangle$, $\langle \Psi_1| P_m| \Psi_1 \rangle \ll 1$. To see this more clearly, we write
\begin{equation}\label{breakPsi}
|\Hat{\Psi}_1 \rangle = c_1|\Psi_1 \rangle + c_2|\Phi \rangle ,
\end{equation}
where $|\Phi \rangle$ is the (orthogonal) deviation from $|\Psi_1 \rangle$. We can then write $\langle \Hat{\Psi}_1|P_m|\Hat{\Psi}_1\rangle = |c_1|^2\langle \Psi_1|P_m|\Psi_1\rangle + R$, with the rest $R = 2\text{Re}(c_1c_2^*\langle \Psi_1|P_m|\Phi\rangle) + |c_2|^2\langle \Phi|P_m|\Phi\rangle$. Assuming $R \ll |c_1|^2\langle \Psi_1|P_m|\Psi_1\rangle$ (since $|c_2| \ll |c_1|$), we can expand the denominator in (\ref{PHat-Def}),
\begin{align}\label{lastArg}
\frac{\text{Tr}(P_m\rho)}{\langle \Hat{\Psi}_1|P_m|\Hat{\Psi}_1\rangle} \approx& \bigg( p_1 + \frac{\sum_{i=2}^np_i\langle \Psi_i|P_m|\Psi_i\rangle}{\langle \Psi_1|P_m|\Psi_1\rangle} \bigg) \nonumber \\
& \times \bigg( 1 - \frac{R}{|c_1|^2 \langle \Psi_1|P_m|\Psi_1\rangle} \bigg)\frac{1}{|c_1|^2} .
\end{align}
We thus find that the $p_1$ estimation attributed to a measurement $P_m$ decreases with growing $R$. This happens if the overlap of $P_m$ with $|\Phi\rangle$ increases. In other words, $p_1$ is underestimated if there exist measurement operators $P_m$ that have sufficiently large overlap with $|\Phi\rangle$ (and hence little overlap with $|\Psi_1\rangle$). This insight will guide us below in choosing the cost function.

Note that there exist proposals to retrieve spectral information of unknown density matrices $\rho$ by applying additional, general random unitaries to $\rho$ \cite{Keyl2001estimating, Enk2012measuring, Alonso2018eigenvalue} (see also Ref.~\cite{Tanaka2014determining}). In contrast, we here use knowledge of the dominant eigenstate to approximate its corresponding eigenvalue.

\section{\label{Sec:Neural_quantum_states} Tomography with neural-network quantum states}

Hereafter, we study how to implement the above introduced scheme for mixed-state reconstruction with the recently developed pure-state tomography based on NQS. To this end, we briefly review the NQS ansatz and its usage for pure state tomography \cite{ Torlai2018neural}.

\subsection*{Neural-network quantum states}

\begin{figure}[htb]
	\includegraphics[width=0.9\columnwidth]{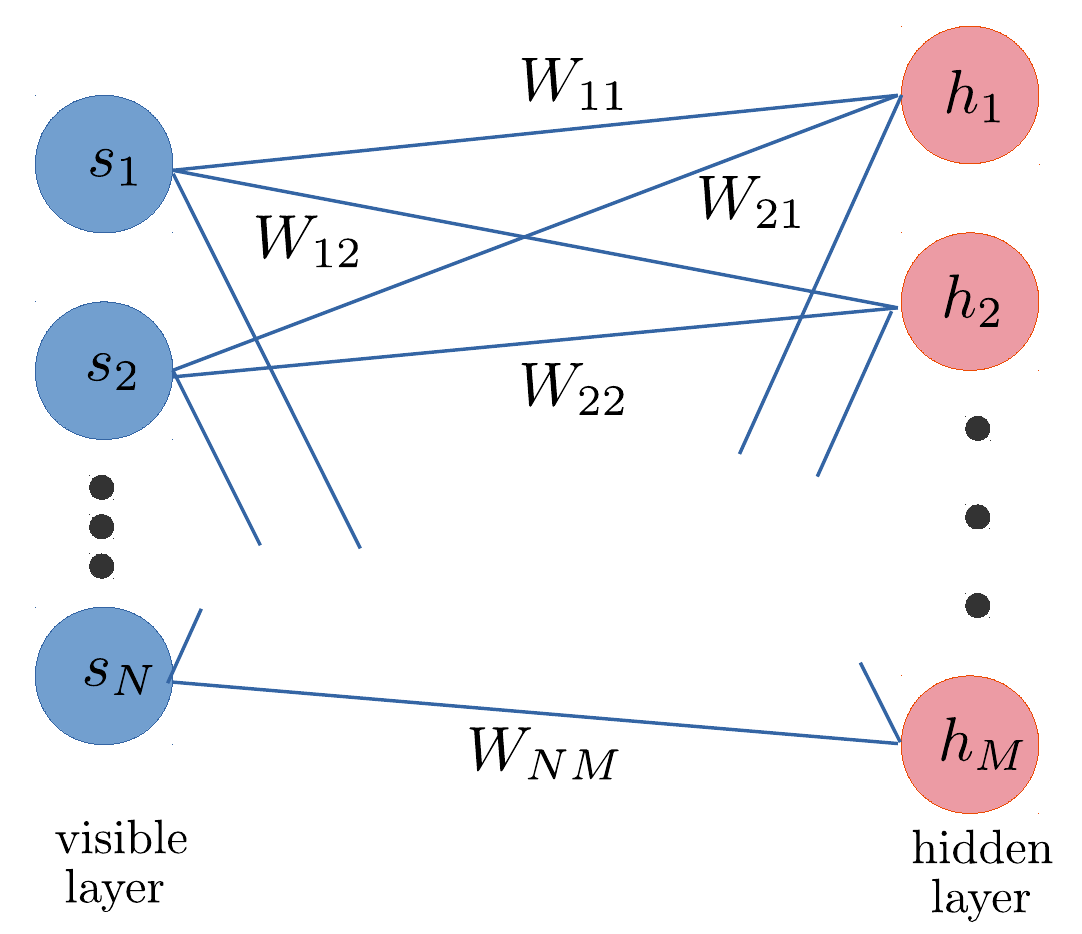}
	\caption{\label{Fig:RBM} Schematic depiction of a restricted Boltzmann machine. Every node $i$ in the visible layer (blue circles) is connected via a weight $W_{ij}$ with every node $j$ in the hidden layer (red circles). There are no intra-layer connections. In addition, all nodes are connected to bias nodes (not depicted). For our applications, the number of hidden nodes can always be chosen equal to the number of visible nodes, $\alpha = N/M = 1$.}
\end{figure}

We consider a quantum system composed of $n$ qubits, with its Hilbert space spanned by some reference basis $\vec{\sigma} = (s_{1}, s_{2}, ..., s_{n})$, with $s_{i} = \pm 1$. A pure quantum state is then completely characterized by the $2^n$ (complex) coefficients $ \langle \vec{\sigma}|\Psi\rangle = \Psi(\vec{\sigma})$. In the definition of the neural-network quantum state ansatz which we implement here \cite{Torlai2018neural}, these coefficients are approximated by two real-valued neural networks, $p_{\lambda}$ and $p_{\mu}$, based on the restricted Boltzmann machine (RBM) architecture, such that
\begin{equation}\label{NQS_Definition}
    \Psi_{\lambda, \mu}(\vec{\sigma}) = \sqrt{\frac{p_{\lambda}(\vec{\sigma})}{Z_{\lambda}}}\exp{[i\Phi_{\mu}(\vec{\sigma})/2]} ,
\end{equation}
where $\Phi_{\mu}(\vec{\sigma}) = \log{p_{\mu}(\vec{\sigma})}$, and $Z_\lambda$ denotes a normalization constant. We remark that alternative neural-network quantum state implementations exist, e.g., based on complex-valued RBMs \cite{Carleo2017solving}.

Briefly, an RBM consists of two layers: the visible layer with $N$ nodes (visible neurons), corresponding to the physical spins; and the hidden layer with $M$ (in our case equal to $N$) auxiliary nodes $h_i$ (hidden neurons). The hidden neurons are coupled to the visible ones, but there is no coupling among neurons in the same layer, as schematically illustrated in Fig.~\ref{Fig:RBM}. Consequently, an RBM can be expressed in the succinct form:
\begin{align}\label{RBM_Definition}
    p_\kappa(\vec{\sigma}, \vec{h}) = \exp{\Big( \sum_{ij}W_{ij}^{\kappa}s_ih_j + \sum_ia_i^\kappa s_i + \sum_jb_j^\kappa h_j \Big)}.
\end{align}
The edge weights $W_{ij}$ and the bias weights $a_i$ and $b_i$ form the parameters to be optimized, subject to some data-based training. If trained successfully, then an RBM delivers a compact approximation to a probability distribution, characterized by the network parameters. Such a compressed representation of the underlying probability distribution then reduces the risk of overfitting \cite{Torlai2019machine}. The distribution of interest can be retrieved by either marginalizing over the hidden states:
\begin{align}\label{Marginalize_RBM}
    p_\kappa(\vec{\sigma}) &= \sum_{\vec{h}}p_\kappa(\vec{\sigma}, \vec{h})\nonumber \\
    &= \exp{\Big( \sum_ia_i^\kappa s_i \Big)}\prod_j 2 \cosh{ \Big( \sum_{i}W_{ij}^{\kappa}s_i +b_j^\kappa \Big)},
\end{align}
or by sampling from the distribution via alternate Gibbs sampling. Sampling is generally efficient, in particular in high dimensions, which contributes to the attractiveness of the RBM model. Moreover, sampling can be used to efficiently calculate the expectation values of many physical observables \cite{Boltzmann_sampling_equilibrium}\cite{qucumber}. We discuss sampling in some detail in Appendix ~\ref{Sec:Gibbs}.

RBMs, and specifically NQS, are steadily gaining popularity in condensed matter and many-body quantum physics, and have already been successfully applied in a wide range of problems, ranging from studying topological states with long-range quantum entanglement \cite{DongLing2017Topological, Gao2017efficient}, to maximizing the violation of Bell inequalities \cite{DongLing2018BellNonLocality}, to determining steady states of dissipative many-body systems \cite{Yoshioka2019constructing, Vicentini2019variational, Hartmann2019neural, Nagy2019variational}. Open-source packages, accelerating and facilitating their implementation, are readily available \cite{qucumber, netket:2019}.

\subsection*{Pure-state tomography}

We now discuss the usage of NQS for the tomography of pure states, as introduced in \cite{Torlai2018neural}. The starting point for the reconstruction is a series of independent projection measurements on a pure state, $|\psi(\vec{\sigma}^{[b]})|^2 (:= P_b(\vec{\sigma}^{[b]}))$. Here the basis rotations $b$ are applied to $\vec{\sigma}$ to obtain a collection of projection bases  $\vec{\sigma}^{[b]}$. The RBMs are then trained on this data set such that the network parameters, $\lambda$ and $\mu$, maximize the data-set likelihood, i.e., $|\Psi_{\lambda, \mu}(\vec{\sigma}^{[b]})|^2 \approx P_b(\vec{\sigma}^{[b]})$. For simplicity (and following Ref.~\cite{Torlai2018neural}), we assume that both RBMs, $p_{\lambda}$ and $p_{\mu}$, feature an equal number of hidden and visible nodes, $N = M$.

The Kullback-Leibler divergence, which quantifies the statistical distance between two probability distributions, can be used as cost function,
\begin{equation}\label{KL1-Definition}
    C = \sum_{b}\text{KL}_{b}^{(1)} = \sum_{b}\sum_{\vec{\sigma}^{[b]}}P_b(\vec{\sigma}^{[b]})\log{\frac{P_b(\vec{\sigma}^{[b]})}{|\Psi_{\lambda, \mu}(\vec{\sigma}^{[b]})|^2}} .
\end{equation}

Typically, the cost is minimized iteratively by gradient descent \cite{Torlai2018neural}, but with increasing system sizes the calculation of the gradients may become intractable. This increase in computational cost is overcome by techniques like stochastic gradient descent and Monte Carlo simulation based on block Gibbs sampling (Appendix~\ref{Sec:Gibbs}). We remark that other choices for cost functions are conceivable and may result in improved performance, e.g., the contrastive divergence between the data and the RBM after a sequence of $k$ block Gibbs sampling steps \cite{Torlai2019machine}.

Pure-state tomography with NQS has been shown to reliably reconstruct pure states of up to 20 qubits, reaching double-nine fidelities \cite{Torlai2018neural, Torlai2019integrating, Torlai2019machine}. Moreover, the method has been shown to be robust under Gaussian noise prevalent in tomographic measurements, to perform well on physically relevant many-body and quantum optics states, and to be efficient by the use of Gibbs sampling on the RBMs.

A possible way to generalize neural-network state tomography to mixed states relies on purification, i.e., the system is augmented by a quantum environment, which subsequently is traced out \cite{Torlai2018latent}. However, this partial trace over the auxiliary degrees of freedom complicates the training procedure unfavorably for large system sizes \cite{carrasquilla2019reconstructing}, and demonstrations have been restricted to small system sizes ($N=2$) \cite{Torlai2018latent}. Here, we take the alternative approach of leveraging powerful neural-network pure-state tomography towards the iterative, eigenstate-wise reconstruction of mixed states.

\subsection*{Mixed-state tomography}\label{CostsSection}

To combine neural-network tomography for pure states with our iterative state-reconstruction scheme, we first need to reexamine the choice of cost function. This may be surprising, since above we have shown that, in ideal conditions, the determination of the dominant eigenstate is robust under different choices of the distance measure. This would then suggest to base the cost on any convenient distance measure, e.g., motivated by Eq.~\eqref{TD_And_Projectors},
\begin{equation}\label{defineL1}
{\rm L}1 = \sum_m|\text{Tr}(P_m\rho)-\langle \hat{\Psi}_1|P_m|\hat{\Psi}_1\rangle| .
\end{equation}

Alternatively, one might wish to use the Kullback-Leibler divergence, as in pure-state tomography.
However, in realistic conditions, for instance, if the data is noisy, different distance measures behave differently, and thus produce results of varying quality. This is because different distance measures tend to emphasize different statistical properties: While statistical distances like L2 or the Kullback-Leibler divergence tend to underscore the importance of larger probability values, as these are statistically most relevant, other choices like L1 put more weight on small probability values.

\begin{figure}[htb]
	\includegraphics[width=\columnwidth]{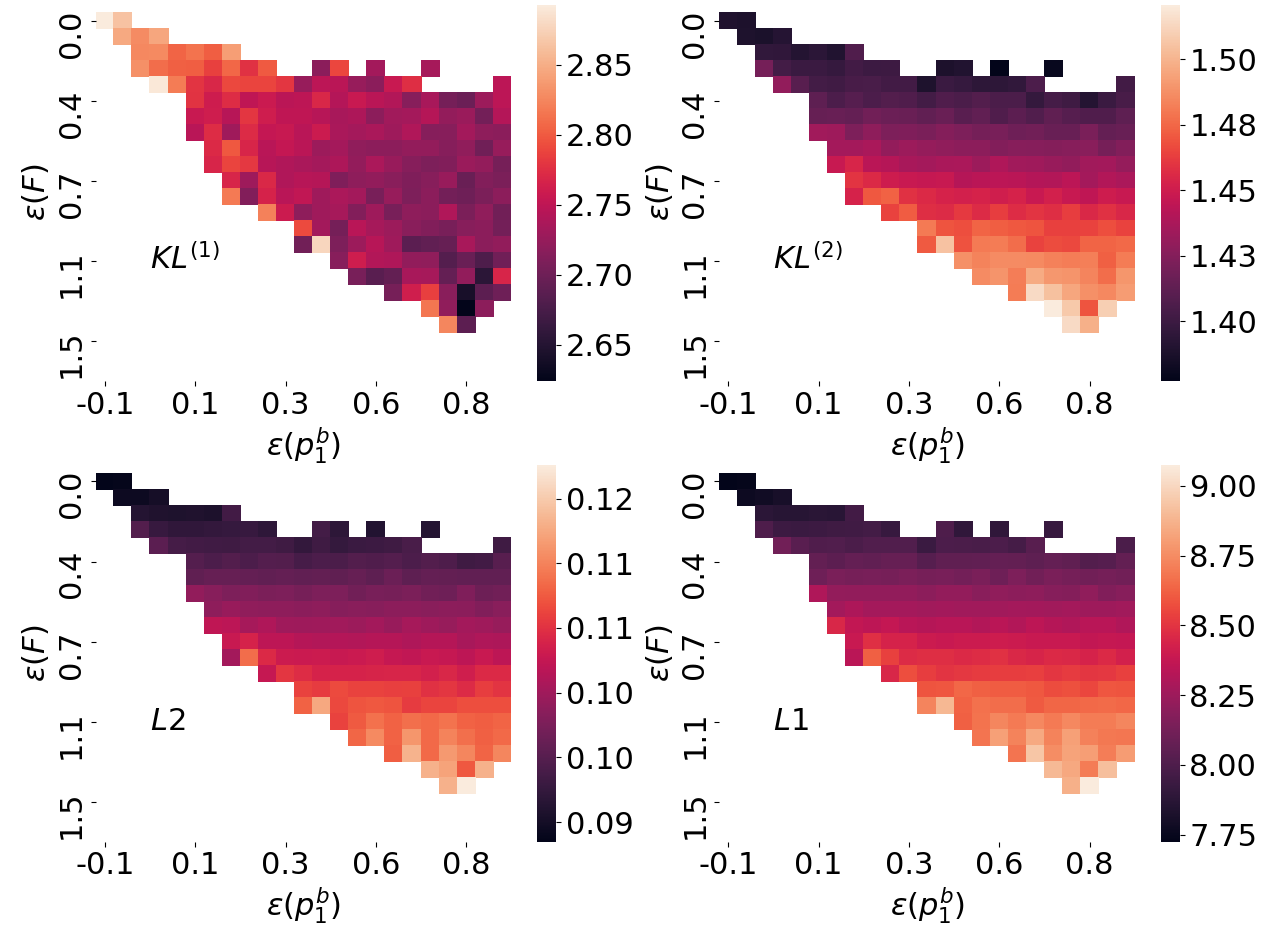}
	\caption[]{\label{Fig:costComparison} Performance of different cost functions in quantifying the statistical difference between measurements on a generic mixed state (taken from a trapped-ions experiment, where 4 ion qubits were prepared in an approximate $W$ state \cite{Haffner2005scalable}) and a pure state. The latter is generated by applying a small random unitary rotation to the true dominant eigenstate. These states are then arranged by their fidelity with respect to the mixed state along the column [$\epsilon(F) = 6000*(1 - \text{Fidelity})$] and by their corresponding value of $p_1^b$ along the row [$\epsilon(p_1^b) = 10*(p_1 - p_1^b)/p_1$].  Here, ${\rm L_k} = \sum_m|\text{Tr}(P_m\rho)-\langle \Psi_1|P_m|\Psi_1\rangle|^{k}$, $\text{KL}^{(1)} = \sum_{m}\text{Tr}(P_m\rho) \log{\frac{\text{Tr}(P_m\rho)}{|\langle \Psi_1|P_m|\Psi_1\rangle|^2}}$, and $\text{KL}^{(2)} = \sum_{m}|\langle \Psi_1|P_m|\Psi_1\rangle|^2 \log{\frac{|\langle \Psi_1|P_m|\Psi_1\rangle|^2}{\text{Tr}(P_m\rho)}}$. Clearly, $\text{KL}^{(1)}$ is most unfaithful to the fidelity.}
\end{figure}

In our case, the distance measure should feature a balanced treatment of the extremes of small and large probabilites: On the one hand, we have seen above that measurements exhibiting small detection probabilities have a significant impact on the quality of the estimation of the dominant eigenvalue $p_1$ [cf.~Eq.~\eqref{lastArg}]; on the other hand, measurements with large detection probabilities are statistically significant for the reliable estimation of the dominant eigenstate $|\Psi_1\rangle$ [cf.~Eq.~\eqref{TD_And_Projectors}]. In numerical experiments, we found that L1.5 performs best for our purpose, cf.~Fig.~\ref{Fig:costComparison}:
\begin{equation}\label{defineL15}
{\rm L}1.5 = \sum_m|\text{Tr}(P_m\rho)-\langle \hat{\Psi}_1|P_m|\hat{\Psi}_1\rangle|^{1.5} .
\end{equation}
In addition, the training of the subsequent, subdominant eigenstates is improved by adding to the cost the overlap with the previously learnt dominating eigenstates, enforcing their orthogonality.

As a proof-of-principle demonstration, we discuss our scheme with a two-qubit state, prepared in a mixture of the four Bell states, with the respective eigenvalues $p_1=0.9$, $p_2=0.09$, $p_3=0.009$, and $p_4=0.001$ (Note that this choice preserves the presence of a dominant eigenstate at any iteration depth). The measurement data is based on the Pauli observables. We obtain excellent recovery of the first (dominant) eigenstate/eigenvalue pair, with $|\langle \Hat{\Psi}_1|\Psi_1\rangle|^2 = 0.99995$ and $p_1^b = 0.9002$. In the second iteration step, we still obtain an excellent eigenstate approximation, $|\langle \Hat{\Psi}_2|\Psi_2\rangle|^2 = 0.99998$; the eigenvalue, however, is underestimated by about $15\%$, $p_2^b = 0.077$, which hints at an error progression from the previous dominant eigenstate approximation, cf.~our discussion in Sec.~\ref{Sec:Methods}. Since this error in the eigenvalue estimation exceeds the tolerance threshold for the next iteration step, we terminate here at an overall reconstruction fidelity of $96.6\%$.

\begin{figure}[htb]
\centering
\includegraphics[width=\columnwidth]{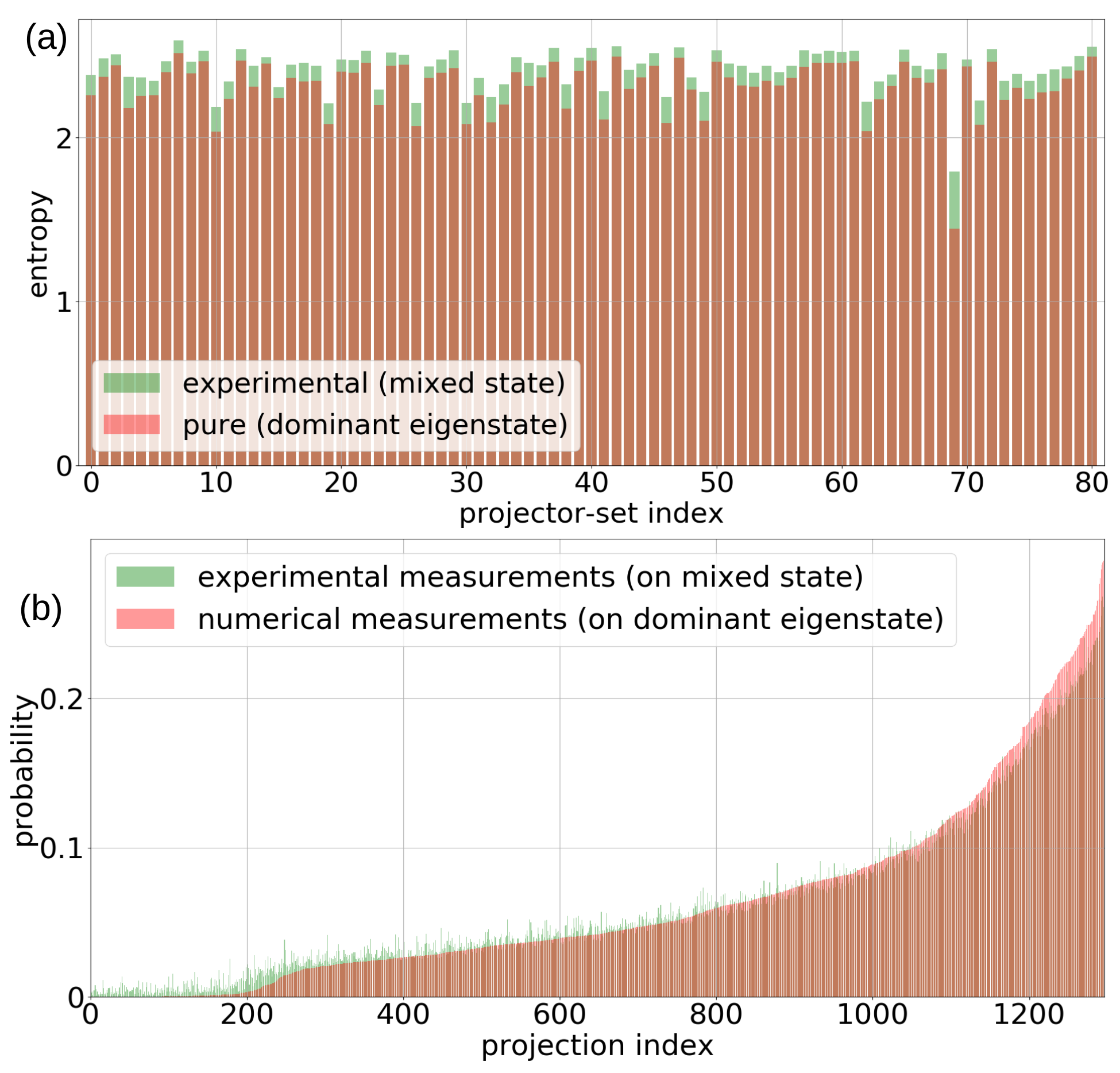}
\caption{\label{Fig:Eigenstate_statistics} The measurement statistics of a mixed state and its dominant eigenstate are strongly correlated and yet exhibit characteristic differences. We demonstrate this with experimental data from a trapped-ions experiment \cite{Haffner2005scalable}, where 4 ion qubits were prepared in an approximate $W$ state. (a) All the $3^4=81$ possible Pauli measurements exhibit reduced entropies for the dominant eigenstate when compared to the full mixed state. (b) Moreover, the dominant eigenstate accentuates measurement outcomes with extreme statistics, enhancing strong and suppressing weak measurement contributions. Shown are the detection probabilities for all the $(2 \times 3)^4=1,296$ possible Pauli-basis projections.}
\end{figure}

Let us remark that we believe that there is still great potential for improving the cost function, by more directly exploiting the characteristic statistical differences between mixed states and their dominant eigenstates. For example, the measurements on the dominant eigenstates exhibit systematically reduced entropies as compared to the full mixed state, cf.~Fig.~\ref{Fig:Eigenstate_statistics}(a). Even more significantly, the eigenstate accentuates measurement outcomes with extreme statistics, enhancing strong, and suppressing weak measurement contributions, see Fig.~\ref{Fig:Eigenstate_statistics}(b). We checked these statistical features for generic random density matrices.  We leave the construction of cost functions that enbody these insights for future work.

To further improve the estimation of $p_1^b$, one may take into account that tomographic measurements come with unavoidable, intrinsic noise, which suggests to discard measurement operators below a preset detection threshold. Moreover, an additional ``noise layer'' in the RBM architecture may further contribute to mitigate measurement errors \cite{Torlai2019integrating}.

\section{\label{Sec:Mixed_state_tomography} Application to Trapped-Ions Experiments}

We evaluated the iterative state reconstruction with neural-network tomography \footnote{Sourcecode available at \href{https://github.com/elMelkani/RBM_mixed_state_tomography/}{https://github.com/elMelkani/ RBM{\textunderscore}mixed{\textunderscore}state{\textunderscore}tomography/}.} using experimental data from trapped-ions experiments \cite{Haffner2005scalable}, where 4 to 8 ion qubits were prepared in approximate $W$ states and subsequently subjected to full state tomography. Referring to compressed sensing \cite{Gross2010quantum}, we constructed the measurement operators $P_m$ from the eigenstates of a random subset of about $3n(\frac{3}{2})^n$ Pauli observables, $\sigma_{i_1} \otimes \sigma_{i_2} \otimes...\otimes\sigma_{i_n}$ ($i \in \{x,y,z\}$), out of the complete set of $3^n$ Pauli observables. We remark that, for the system sizes considered,  the normalized distributions $p_{\lambda}(\vec{\sigma})$ and $p_{\mu}(\vec{\sigma})$ could be determined explicitly and Gibbs sampling was not required.

We applied iterative state reconstruction up to the second eigenstate (rank-$2$ approximation) for system sizes of 4 to 8 qubits. The results are given in Table~\ref{Table:Iterative_reconstruction}. We find that the dominant eigenstate is recovered with $>99\%$ fidelity for all the system sizes considered. This robust performance in the reconstruction of the first eigenstate with above $99\%$ fidelity was also confirmed by additional testing on a family of randomly generated density matrices. The reliable knowledge of the dominant eigenstates then lets us assess the quality of the target state production (e.g., offsets from the (pure) target state caused by systematic errors in the coherent control), in terms of the overlap of the dominant eigenstate with the target $W$ state. This provides an alternative to the more standard assessment in terms of the overlap of the produced mixed state with the target state \cite{Haffner2005scalable}, which contains also contributions from all subdominant eigenstates.

The estimation of the dominant eigenvalue, which informs us about the ``purity offset'' from the (pure) target state (e.g., induced by coupling to an environment, or by parameter drifts between different runs of the experiment \cite{Gneiting2018disorder}), shows a deteriorating scaling behavior: While the dominant eigenvalue is less than $4\%$ off in the 4-qubit case, it is underestimated by about $33\%$ in the case of 8 qubits, which hints at an increasing influence of the error in the eigenstate approximation, cf. our discussion in Section~\ref{Sec:Methods}. While a growing estimation error appears natural with regard to the decreasing purity of the dominant eigenstate contribution, its magnitude may appear surprising in view of the excellent eigenstate approximations; however, as indicated by Eq.~(\ref{lastArg}), measurement projectors with very small overlap with the eigenstate tend to strongly amplify the error of the eigenstate approximation, resulting in erroneous minima in the estimate (\ref{PHat-Def}). Note that this issue could be circumvented in an adaptive measurement scheme, where knowledge of the dominant eigenstates is used to implement measurements that maximize the overlap with the eigenstates.

Along with the error scaling of the estimation of the first eigenvalue, we observe an increasingly poor reconstruction of the second eigenstate and eigenvalue, cf.~Table~\ref{Table:Iterative_reconstruction}. This may be due to the incomplete subtraction of the first eigenstate, cf.~Eq.~(\ref{rho_Dash_Def}), or because, with increasing Hilbert space dimension, the second and the third eigenvalue become comparable of size and the respective eigenstates thus harder to discriminate.

\begin{table*}[htb]
	\begin{tabular}{c| c c c c| c c c c c c |c c} 
		\hline
		 \phantom{a} N \phantom{a} & \phantom{a} $p_1$ \phantom{a} & \phantom{a} $p_2$ \phantom{a} & $\kappa(2)$ & \phantom{a} $p_3$ \phantom{a} & \phantom{a} $|\langle \Hat{\Psi}_1|\Psi_1\rangle|^2$ \phantom{a} & $p_1^b$ &  \phantom{a} $|\langle \Hat{\Psi}_2|\Psi_2\rangle|^2$ \phantom{a} & $p_2^b$ &  \phantom{a} F($\rho, \sigma$) \phantom{a} & $RF(\rho, \sigma)$  &  \phantom{a} $F(\rho, W)$ &  \phantom{a} $|\langle \Hat{\Psi}_1|W\rangle|^2$\\ [0.5ex]
		\hline
		4 & 0.860 & 0.063 & 0.922 & 0.037 & 0.999 & 0.836 & 0.852 & 0.018 & 0.905  & 0.981 & 0.85 & 0.985\\ 
		5 & 0.824 & 0.073 & 0.896 & 0.042 & 0.998 & 0.765 & 0.769 & 0.008 & 0.860 & 0.960 & 0.76 & 0.930\\
		6 & 0.813 & 0.070 & 0.883 & 0.042 & 0.998 & 0.690 & 0.801 & 0.010 & 0.865  & 0.979 & 0.79 & 0.974\\
		7 & 0.782 & 0.060 & 0.843 & 0.044 & 0.993 & 0.545 & 0.284 & 0.008 & 0.805  & 0.955 & 0.76 & 0.981\\
		8 & 0.751 & 0.061 & 0.812 & 0.046 & 0.994 & 0.505 & 0.246 & $\approx$ 0 & 0.748  & 0.922 & 0.72 & 0.959\\
		\hline
	\end{tabular}
	\caption{\label{Table:Iterative_reconstruction} Iterative rank-$2$ reconstruction of the approximate $W$ states produced in \cite{Haffner2005scalable}, ranging from 4 to 8 ion qubits. The four left columns specify the state $\rho$ in terms of its leading eigenvalues, with $\kappa(2) := p_1 + p_2$ being the maximum fidelity a rank-$2$ approximation of $\rho$ can achieve. The center six columns display the outcome of the iterative reconstruction. The overlap $|\langle \Hat{\Psi}_1|\Psi_1\rangle|^2$ between the dominant eigenstate $|\Psi_1\rangle$ and its approximation $|\Hat{\Psi}_1\rangle$ learned by the NQS maintains double-9 quality through all system sizes considered. The reconstructed second eigenstate and the estimates $p_i^b$ of the respective eigenvalues $p_i$ display deteriorating scaling behavior. The rank-$2$ approximations $\sigma := \hat{\kappa}(2)^{-1}\sum_{i=1}^2 p_i^b|\Hat{\Psi}_i\rangle\langle\Hat{\Psi}_i|$, with $\hat{\kappa}(2) = p_1^b + p_2^b$ for normalization, feature fidelities $F(\rho, \sigma)$ with the actual mixed states $\rho$ ranging from 0.90 (4 qubits) to 0.74 (8 qubits). This results in relative fidelities $RF(\rho, \sigma) = F(\rho, \sigma)/\kappa(2)$ ranging from 0.98 (4 qubits) to 0.92 (8 qubits). The rightmost two columns display the fidelity of $\rho$ with the target $W$ state (cf.~\cite{Haffner2005scalable}), and the fidelity of the approximated dominant eigenstate $|\Hat{\Psi}_1\rangle$ with the target $W$ state. Our results confirm that the recovered dominant eigenstate/eigenvalue pair provides us with a viable assessment of the quality of the target state production.}
\end{table*}

We define the relative fidelity $RF(\rho, \sigma)$ as the ratio between the fidelity achieved and the maximum fidelity possible at a given rank $r$, $RF(\rho, \sigma) = F(\rho, \sigma)/\kappa(r)$. We find that the relative fidelity drops from 0.98 at 4 qubits to (still competitive) 0.92 at 8 qubits. For comparison, Ref.~\cite{Gross2010quantum} reports a rank-$3$ reconstruction fidelity of 0.82 for the same approximate 8 qubit $W$ state, which translates into a relative fidelity of 0.95. (Ref.~\cite{Gross2010quantum} reports a rank-$3$ reconstruction fidelity of 0.91, which is, however, larger than the maximum fidelity achievable in our definition of fidelity, $\kappa(3) = 0.86$. We thus assume they chose the square-root convention to define the fidelity.)

We still see large potential for an improved scaling behavior in the reconstruction of the state properties beyond the dominant eigenstate. This may be achieved, for example, by estimating both eigenstates and eigenvalues simultaneously, by employing a three-layer generalization of NQS \cite{Symmetries_excited}, or by implementing adaptive measurement schemes. Irrespectively, the dominant eigenvalue estimate may be taken as a fair assessment of the overall quality of the state produced, which, along with the dominant eigenstate, provides the arguably most relevant information about the state production.

We remark that we also tested the quality of the eigenstate recovery with training data significantly reduced below the compressed sensing threshold. In that case, the training data is presumably not sufficient to single out a unique mixed state. Nevertheless, we still obtained very good agreement for the dominant eigenstate approximation. This may hint at the greater robustness of pure-state tomography, and at the supportive generalization behavior of the neural-network ansatz. In the case of incomplete measurement data, one may then test against overfitting by splitting the data into a training and a test set.

\section{Conclusions}

We presented a scheme for quantum state tomography via the stepwise retrieval of the eigenstates and eigenvalues of the mixed states produced in experiments. Our scheme iteratively exploits that dominant eigenstates can be robustly extracted from mixed-state measurement data using pure-state tomography methods, inheriting their scaling behavior. As a specific method for pure-state tomography, we chose the efficient and scalable representation and training of pure states based on restricted Boltzmann machines. We demonstrated our scheme with experimental data from trapped-ions experiments, where approximate $W$ states from 4 to 8 qubits were produced. We find that the dominant eigenstates can be excellently retrieved, with fidelities consistently exceeding 0.99. In the 4-qubit case, we reach an overall fidelity of 0.90 for a rank-$2$ approximation, which corresponds to a relative fidelity of 0.98. In the 8 qubit case, we still reach an overall fidelity of 0.75 for a rank-$2$ approximation, corresponding to a relative fidelity of 0.92.

Our scheme is designed to deliver low-rank approximations, following the cost scaling of pure-state tomography and with the rank not required to be set {\it a priori}. It is particularly well-suited for the experimentally relevant case of density matrices exhibiting dominant eigenstates, where pure-state tomography methods can be expected to produce accurate approximations of the latter. In contrast to a full matrix reconstruction, which scales as $\mathcal{O}(n^2)$ in computational cost, our procedure scales as $\mathcal{O}(n r)$, where $r$ denotes the rank to be achieved. Moreover, the computational burden is mitigated, since in each step only a pure state needs to be processed, which can then be stored separately. We conjecture that the stepwise optimization of pure states may, due to their inherent coherence (which constrains the measurement statistics) and in line with compressed sensing, also have a positive effect on the required amount of training data.

From a conceptual perspective, our scheme directly and efficiently delivers the arguably most relevant information about successful state production in experiments: The leading eigenstate(s) can inform the experimenter about systematic errors in the state production, while the dominant eigenvalue captures and quantifies the impact of decoherence and uncontrolled parameter fluctuations. This reasoning also straightforwardly generalizes to cases where the target states are (low-rank) mixed states. The practicality of such condensed and structured assessment of experimentally produced states can only increase with growing system size and exponentially growing Hilbert space dimension.

We still see considerable potential for improving the eigenvalue estimation. Our present method displays a consistent underestimation of the eigenvalues, with an error of less than $4\%$ in the 4-qubit case, which rises to about $33\%$ in the case of 8 qubits. Possible improvement strategies include modified neural-network architectures, more refined eigenvalue estimations, cost functions that are optimized with respect to the statistical characteristics of eigenstates, and adaptive measurement schemes. We leave this for future research.

\paragraph*{Acknowledgments.}
We thank Daoyi Dong for helpful comments on the manuscript.
F.N. is supported in part by:
NTT Research,
Army Research Office (ARO) (Grant No.~W911NF-18-1-0358),
Japan Science and Technology Agency (JST) (via the Q-LEAP program and the CREST Grant No.~JPMJCR1676),
Japan Society for the Promotion of Science (JSPS) (via the KAKENHI Grant No.~JP20H00134, and the
JSPS-RFBR Grant No.~JPJSBP120194828),
and the Foundational Questions Institute Fund (FQXi) (Grant No.~FQXi-IAF19-06), a donor advised fund of the Silicon Valley Community Foundation.

\bibliography{references}

\appendix

\section{Gibbs sampling \label{Sec:Gibbs}}

Gibbs sampling of a restricted Boltzmann machine (RBM) enables one to infer the values of the probability distribution $P(\vec{\sigma}) = p(\vec{\sigma})/Z$ without the need to calculate every $p(\vec{\sigma})$ individually, as it would be required to determine the normalization $Z$. To this end, one devises a two-step Markov-chain sampling with transition probabilities:
\begin{equation}\label{Gibbs1}
    T_\sigma[(\vec{\sigma}, \vec{h}) \xrightarrow{} (\vec{\sigma}', \vec{h})] = \frac{P(\vec{\sigma}', \vec{h})}{\sum_{\vec{\sigma}''}P(\vec{\sigma}'', \vec{h})} = P(\vec{\sigma}'| \vec{h}) ,
\end{equation}
\begin{equation}\label{Gibbs2}
    T_\sigma[(\vec{\sigma}, \vec{h}) \xrightarrow{} (\vec{\sigma}, \vec{h}')] = \frac{P(\vec{\sigma}, \vec{h}')}{\sum_{\vec{h}''}P(\vec{\sigma}, \vec{h}'')} = P(\vec{h}'|\vec{\sigma}) .
\end{equation}
That is, from a given configuration $(\vec{\sigma}, \vec{h})$ of the RBM, we can treat either the visible nodes $\vec{\sigma}$ as evolving stochastically as a function of the hidden nodes $\vec{h}$ or vice versa. These conditional probabilities are very efficient to compute, as nodes within same layers are independent of each other. For example,
\begin{align}\label{Gibbs3}
    P(\vec{\sigma}'| \vec{h}) &= \frac{ \exp{ \big( \sum_{ij}W_{ij}^{\kappa}s'_ih_j + \sum_ia_i^\kappa s'_i + \sum_j b_j^\kappa h_j} \big)} {\sum_{\vec{\sigma}''}e^{\sum_{ij}W_{ij}^{\kappa}s''_ih_j + \sum_ia_i^\kappa s''_i + \sum_jb_j^\kappa h_j}} \nonumber \\
    &= \frac{ \prod_i \exp{\big( \sum_{j}W_{ij}^{\kappa}s'_ih_j + a_i^\kappa s'_i \big)}}{\prod_i \Big( e^{\sum_{j}W_{ij}^{\kappa}h_j + a_i^\kappa} + e^{-\sum_{j}W_{ij}^{\kappa}h_j - a_i^\kappa} \Big)} .
\end{align}
We can factorize this expression to obtain
\begin{equation}\label{Gibbs4}
    P(s_i' = 1| \vec{h}) = \frac{1}{1 + \exp{[-2(\sum_{j}W_{ij}^{\kappa}h_j + a_i^\kappa)]}} ,
\end{equation}
and similarly,
\begin{equation}\label{Gibbs5}
    P(h_j' = 1| \vec{\sigma}) = \frac{1}{1 + \exp{[-2(\sum_{i}W_{ij}^{\kappa}s_i + b_j^\kappa)]}} .
\end{equation}
Repeating these Markov-chain steps $N_s$ times from a randomly generated initial configuration, the resulting $\vec{\sigma}$ is effectively sampled from the distribution $P(\vec{\sigma})$, regardless of the starting point. Producing sufficiently many samples according to the distribution then allows to infer $p(\vec{\sigma})/Z$, as it is needed to determine the gradients.

This provides a method to circumvent the problem of needing to calculate explicitly all the outputs $p_{\lambda}(\vec{\sigma}_i)$ and $p_{\mu}(\vec{\sigma}_i)$ corresponding to each input $\vec{\sigma}_i$ which becomes very expensive when the state space enlarges to, say, order of $2^{20}$.

\section{Proofs of propositions \label{Sec:App-Theory}}

In the following we present the proofs of the propositions from Section~\ref{Sec:Theoretical_Results}.

\subsection{Rank-$1$ approximations}

The case of rank-$1$ approximations is special for two reasons: First, the rank-$1$ approximation of a density matrix corresponds to its closest pure state and hence is conceptually important. Second, in this case both fidelity and trace distance are optimized uniquely by the same pure state.

\subsubsection{Fidelity}
The unique closest pure state to $\rho$ is its dominant eigenstate $|\Psi_1\rangle$.

The proof is trivial:
Let $|\Phi\rangle$ be the closest pure state. Note that $\tau = |\Phi\rangle\langle\Phi|$ has eigenvalues $1$ (with multiplicity $1$) and $0$ (with multiplicity $n - 1$ ), so that $\sqrt{\tau} = \tau$. Then,
\begin{align}\label{Fide1-Proof}
F(\rho, \tau = |\Phi\rangle\langle\Phi|) \nonumber &= \bigg [\text{Tr}\bigg (\sqrt{\sqrt{\tau} \rho \sqrt{\tau}}\bigg )\bigg ]^2 \nonumber \\
&= \bigg [\sqrt{\langle\Phi| \rho |\Phi\rangle}\bigg ]^2 \nonumber \\
&= \langle\Phi| \rho |\Phi\rangle ,
\end{align}
which is maximized uniquely by $|\Phi\rangle = |\Psi_1\rangle$, with fidelity $p_1$.

\subsubsection{Trace distance}

We claim that the minimum trace distance $T$ between $\rho$ and a pure state is bounded by $1 - p_1$. A straightforward computation then shows that $|\Psi_1\rangle$ reaches this bound.

To prove the claim, and that $|\Psi_1\rangle$ is the unique solution, we make use of Weyl's inequality \cite{Weyl}: Let $Q$ and $P$ be two Hermitian matrices, and define $M = Q + P$. Label the eigenvalues of $Q$ by $q_i$, of $P$ by $p_i$, and of $M$ by $m_i$, such that $q_1 \geq q_2 \geq ... \geq q_n$, $p_1 \geq p_2 \geq ... \geq p_n$, and $m_1 \geq m_2 \geq ... \geq m_n$. Weyl's inequality then states that
\begin{equation}\label{Weyl}
    q_j + p_k \leq m_i \leq q_r + p_s
\end{equation}
whenever
\begin{equation}\label{WCondition}
    j + k - n \geq i \geq r + s - 1 .
\end{equation}
In our case $M = Q + P = (-\sigma) + \rho $, and therefore $(q_1, q_2, ..., q_{n-1}, q_n) = (0, 0, ..., 0, -1)$, as $\sigma$ is a pure state. First, let $j, r, i, k = n$ and $s = 1$ (one can check that condition \eqref{WCondition} holds), so that Weyl's inequality \eqref{Weyl} gives:
\begin{align}\label{res1}
    p_n - 1 \leq m_n \leq p_1 - 1 \nonumber \\ 
    \implies 1 - p_1 \leq |m_n| \leq 1 - p_n
\end{align}
Second, let $r = 1, j = n - 1, s = i, k = i +1$ for $i = 1$ to $n-1$, so that Weyl's inequality \eqref{Weyl} gives:
\begin{align}\label{res2}
    p_{i+1} \leq m_i \leq p_i \nonumber \\
    \implies p_{i+1} \leq |m_i| \leq p_i
\end{align}
for $i = 1$ to $n-1$. Therefore, using Eqs.~\eqref{res1} and \eqref{res2},
\begin{align*}
    (1 - p_1) + \sum_{i=1}^n p_{i+1} &\leq \sum_{i=1}^n |m_i| \leq (1 - p_n) + \sum_{i=1}^n p_{i}\\
    2(1 - p_1) &\leq \sum_{i=1}^n |m_i| \leq 2(1 - p_n) ,
\end{align*}
or
\begin{equation}\label{finalRes}
    (1 - p_1) \leq T(\rho, \sigma) = \frac{1}{2}\sum_{i=1}^n|m_i| \leq (1 - p_n) .
\end{equation}
The inequality, $(1 - p_1) \leq T(\rho, \sigma)$ becomes an equality if $m_n = p_1 + q_n = p_1 - 1$ and $p_{i+1} + q_{n-1} = p_{i+1} = m_i$. For this to happen it is necessary that the eigenspaces corresponding to the eigenvalues in the equations have nonvanishing intersection \cite{HORN}. In particular, it is required that the eigenspace corresponding to $p_1$, which is given by ${|\Psi_1\rangle}$, and the eigenspace corresponding to $q_n$, which is given by ${|\Phi\rangle}$, have a nonvanishing intersection, implying that $|\Phi\rangle = |\Psi_1\rangle$, uniquely.

\subsection{Rank-$r$ approximations}

\subsubsection{Fidelity}

Let $\tau = \sum_{i=1}^r q_i |\Phi_i\rangle\langle\Phi_i|$ be the closest rank-$r$ approximation with $|\Phi_i\rangle \in \boldsymbol{H}^{(r)}$ and $\boldsymbol{H}^{(r)}$ an $r$-dimensional (Hilbert-)subspace of $\boldsymbol{H}^{(n)}$. We will work in the basis of the solution, so that $\tau = \text{diag} (q_1, q_2, ..., q_r, 0, ..., 0)$. Define $D = \sum_{i=1}^r |\Phi_i\rangle\langle\Phi_i|$ and consider the normalized $r\times r$ submatrix of $\rho$ given by: 
\begin{equation}\label{truncated_Rho}
    \rho' = \frac{D\rho D}{\text{Tr}(D\rho D)} .
\end{equation}
By construction, $\rho'$ is Hermitian and has trace one. To verify that it is also positive semi-definite, we argue that, for any $|x\rangle$ in the Hilbert space (ignoring the normalization term for $\rho'$, which is positive),
\begin{align}\label{positivity}
    \langle x| \rho' |x\rangle &= \langle x| D\rho D |x\rangle \nonumber\\
    &= \frac{\langle x| D}{\langle x| D^{\dagger} D |x\rangle} \rho \frac{ D |x\rangle}{\langle x| D^{\dagger} D |x\rangle} \langle x| D^{\dagger} D |x\rangle^2 \nonumber \\
    &= \langle x'|\rho |x'\rangle \langle x| D |x\rangle^2 ,
\end{align}
where $|x'\rangle = \frac{ D |x\rangle}{\langle x| D^{\dagger} D |x\rangle}$ also lies in the Hilbert space, and we used $D^{\dagger} D = D$. Noting that both of these terms in the multiplication are nonnegative, we conclude that $\rho '$ is nonnegative, i.e., positive semi-definite. Hence, since $\rho '$ is Hermitian, has trace one, and is positive semi-definite, it constitutes a valid density matrix (of at most rank $r$). For the fidelity we then obtain
\begin{align} \label{FideStep2}
F(\rho, \tau) &= \bigg [\text{Tr}\bigg (\sqrt{\sqrt{\tau} \rho \sqrt{\tau}}\bigg )\bigg ]^2 \nonumber \\
&= \bigg [\text{Tr}\bigg (\sqrt{\sqrt{\tau} \rho ' \sqrt{\tau}}\bigg )\bigg ]^2 \text{Tr}(D\rho D) \nonumber \\
&= F(\rho ', \tau) \text{Tr}(D\rho D) ,
\end{align}
where we have used $ D \sqrt{\tau} = \sqrt{\tau} D = \sqrt{\tau} $. Note that the first term is maximized (to one) when both rank-$r$ matrices are the same, that is, $\tau = \rho'$. The second term, on the other hand, simplifies to
\begin{align}\label{FideProof3}
\text{Tr}(D\rho D)&=\text{Tr} \Big(\sum_{i=1,j=1,k=1}^{i=r,j=n,k=r}|\Phi_i\rangle \langle \Phi_i| p_j |\Psi_j\rangle \langle \Psi_j|\Phi_k\rangle \langle \Phi_k| \Big) \nonumber \\
&=\sum_{i=1,j=1}^{i=r,j=n} p_j |\langle \Phi_i|\Psi_j\rangle|^2 .
\end{align}
Now let $\sum_{i=1}^r |\langle \Phi_i|\Psi_j\rangle|^2 = k_j \leq 1$, where the inequality becomes an equality if  $|\Psi_j\rangle$ can be decomposed into the incomplete $|\Phi_i\rangle$ basis. We can then write
\begin{equation}\label{lastStep}
    \text{Tr}(D\rho D) = \sum_{j=1}^n p_j k_j ,
\end{equation}
which is maximized (to $\kappa$) when $k_j = 1$ for $1 \leq j \leq r$ (then, $k_j = 0$ for $r+1 \leq j \leq n$). In other words, the first $r$ dominant eigenvectors $|\Psi_j\rangle$ lie in the same subspace as spanned by the $|\Phi_i\rangle$. Maximization follows, since any other choice of $k_j$ would have to trade the factor of a large $p_i$ (small $i$) for a larger factor accompanying a small $p_i$ (large $i$), which would reduce the trace. Consequently, $\tau = \rho'$ is the unique solution maximizing both terms. Since $\rho$ is diagonal in that case, we obtain $\rho' = \sigma$. Therefore, $\sigma$ is the unique rank-$r$ approximation of $\rho$ that is closest to it (in terms of fidelity), with fidelity $F(\rho, \sigma) = \kappa$.

\subsubsection{Trace Distance}

We conjecture, on the basis of numerical experiments, that $\sigma$ also minimizes the trace distance to $\rho$, with $T(\rho, \sigma) = 1 - \kappa$. Here, however, we show that there exist infinitely many rank-$r$ approximations which reach this optimization, all having the first $r$ dominant eigenstates of $\rho$ as their support.

Consider a rank-$r$ approximation $\tau = \sum_{i=1}^r q_i |\Psi_i\rangle\langle\Psi_i| = \text{diag} (q_1, q_2, ..., q_r, 0, ..., 0)$, such that $q_i \geq p_i$ for $i=1$ to $r$. Since $\sum_{i=1}^r q_i = 1 \geq \sum_{i=1}^r p_i$, this condition is satisfied by infinitely many matrices, and $\sigma$ is one of them. The trace distance then evaluates as
\begin{align}\label{TraceProof}
T(\rho, \tau) &= \frac{1}{2} \Big(\sum_{i=1}^r |p_i-q_i| + \sum_{i=r+1}^n |p_i - 0| \Big)\nonumber \\
&= \frac{1}{2} \Big(\sum_{i=1}^r (q_i-p_i) + (1 - \kappa) \Big)\nonumber \\
&= \frac{1}{2} ((1 - \kappa) + (1 - \kappa)) \nonumber \\
&= 1 - \kappa .
\end{align}

\end{document}